\documentclass[twocolumn]{aastex63}

\shorttitle{Stellar surface inhomogeneities in the K2-18~b transmission spectrum}

\graphicspath{{./}{figures/}}

\begin{document}

\title{Stellar surface inhomogeneities as a potential source of the atmospheric signal detected in the K2-18~b transmission spectrum}


\correspondingauthor{Thomas Barclay}
\email{thomas.barclay@nasa.gov}

\author[0000-0001-7139-2724]{Thomas Barclay}
\affiliation{University of Maryland, Baltimore County, 1000 Hilltop Cir, Baltimore, MD 21250, USA}
\affiliation{NASA Goddard Space Flight Center, 8800 Greenbelt Rd, Greenbelt, MD 20771, USA}

\author[0000-0001-9786-1031]{Veselin B. Kostov}
\affiliation{NASA Goddard Space Flight Center, 8800 Greenbelt Rd, Greenbelt, MD 20771, USA}
\affiliation{SETI Institute, 189 Bernardo Avenue, Suite 200, Mountain View, CA 94043, USA}

\author[0000-0001-8020-7121]{Knicole D. Col\'{o}n}
\affiliation{NASA Goddard Space Flight Center, 8800 Greenbelt Rd, Greenbelt, MD 20771, USA}

\author[0000-0003-1309-2904]{Elisa V. Quintana}
\affiliation{NASA Goddard Space Flight Center, 8800 Greenbelt Rd, Greenbelt, MD 20771, USA}

\author[0000-0001-5347-7062]{Joshua E. Schlieder}
\affiliation{NASA Goddard Space Flight Center, 8800 Greenbelt Rd, Greenbelt, MD 20771, USA}

\author[0000-0002-2457-272X]{Dana R. Louie}
\affiliation{NASA Goddard Space Flight Center, 8800 Greenbelt Rd, Greenbelt, MD 20771, USA}

\author[0000-0002-0388-8004]{Emily A. Gilbert}
\affiliation{Department of Astronomy and Astrophysics, University of
Chicago, 5640 S. Ellis Ave, Chicago, IL 60637, USA}
\affiliation{University of Maryland, Baltimore County, 1000 Hilltop Circle, Baltimore, MD 21250, USA}

\author[0000-0001-7106-4683]{Susan E. Mullally}
\affiliation{Space Telescope Science Institute, 3700 San Martin Drive, Baltimore, MD 21218, USA}

\begin{abstract}

Transmission spectroscopy of transiting exoplanets is a proven technique that can yield information on the composition and structure of a planet's atmosphere. However, transmission spectra may be compromised by inhomogeneities in the stellar photosphere. The sub-Neptune-sized habitable zone planet K2-18 b has water absorption detected in its atmosphere using data from the Hubble Space Telescope (HST). Herein, we examine whether the reported planetary atmospheric signal seen from HST transmission spectroscopy of K2-18 b could instead be induced by time-varying star spots. We built a time-variable spectral model of K2-18 that is designed to match the variability amplitude seen in K2 photometric data, and used this model to simulate 1000 HST data-sets that follow the K2-18 b observation strategy. More than 1\% of these provide a better fit to the data than the best-fitting exoplanet atmosphere model. After resampling our simulations to generate synthetic HST observations, we find that 40\% of random draws would produce an atmospheric detection at a level at least as significant as that seen in the actual HST data of K2-18 b. This work illustrates that the inferred detection of an atmosphere on K2-18 b may alternatively be explained by stellar spectral contamination due to the inhomogeneous photosphere of K2-18. We do not rule out a detection of water in the planet's atmosphere, but provide a plausible alternative that should be considered, and conclude that more observations are needed to fully rule out stellar contamination.

\end{abstract}


\section{Introduction} \label{sec:intro}
One of the most exciting discoveries of recent years is that exoplanets are common -- there are likely more planets in the Milky Way than stars \citep{Burke2015,Bryson2020} -- and small planets are frequent around M-dwarfs \citep{Dressing2015,HardegreeUllman2019,Bryson2020b}. 
The next step, beyond understanding coarse demographics, is to perform detailed characterization of individual planets to probe their atmospheres and bulk compositions. Great strides have been made in observing exoplanet atmospheres using data from the Hubble Space Telescope (HST) and Spitzer Space Telescope \citep[e.g.][]{Knutson2011,Sing2016} as well as with ground-based observatories \citep[e.g.][]{Bean2010,Crossfield2013,Stevenson2014,DiamondLowe2018,DiamondLowe2020}. While the majority of atmospheric characterization targets have been giant planets with short orbital periods, a handful of planets smaller than four Earth-radii have been studied \citep[e.g][]{BertaThompson2012,Kreidberg2014,Knutson2014,deWit2016,Crossfield2017,DiamondLowe2018,DiamondLowe2020,Kreidberg2019,Tsiaras2019,Benneke2019,Swain2021}. 

The most prolific method for probing exoplanet atmospheres is transmission spectroscopy \citep{Seager2000,Brown2001,Hubbard2001}. The size of a planet transiting its host star can be determined from the change in a star's brightness during a transit. If an exoplanet has an atmosphere, the opacity can be wavelength dependent, and this will cause a planet to appear larger -- and have a deeper transit -- at wavelengths with atomic and molecular absorbers. Transmission spectroscopy has been used to infer H$_2$O, CO, CO$_2$, HCN, and CH$_4$ abundances \citep[e.g.][]{Kreidberg2014b,MacDonald2017,Wakeford2018}, atmospheric metallicity \citep{Welbanks2019,Colon2020}, cloud and haze coverage \citep{Sing2016,wakeford2019rnaas}, and planet formation mechanisms \citep{Madhusudhan2012,Madhusudhan2014}. 

Characterization of the atmospheres of small planets via transmission spectroscopy is generally only accessible to current generation instruments if they orbit small stars, because the transit depth scales as the square of planet-to-star radius ratio. NASA's Transiting Exoplanet Survey Satellite \citep[TESS, ][]{Ricker2015} is designed to provide a plethora of planets orbiting bright and nearby cool stars \citep{Barclay2018,Kempton2018,Dalba2019} for atmospheric characterization by the Hubble Space Telescope (HST) and, looking further ahead, by the James Webb Space Telescope \citep[JWST,][]{Gardner2006} and the Atmospheric Remote-sensing Infrared Exoplanet Large-survey \citep[Ariel,][]{Tinetti2018}. While JWST provides a significant technical advancement over current instruments, most of the super-Earth and smaller planets accessible to JWST will orbit M-dwarfs\footnote{Over 60\% of JWST Cycle 1 General Observer Exoplanet targets smaller than 3R$_\oplus$ orbit M-dwarfs.}. Likewise, Ariel, which is scheduled to launch in the late 2020s \citep{Tinetti2018}, will be most sensitive to the atmospheres of small planets if they orbit small stars \citep{Edwards2019}.

However, small stars are frequently magnetically active \citep{Shields2016}. The level of activity can be estimated by observing flares \citep{gunther2019}, high UV fluxes \citep{Melbourne2020}, and periodic changes in brightness associated with surface inhomogeneities \citep{Raetz2020}. These inhomogeneities include dark spots and bright faculae. Spots can cover 10--50\% of the surface of a cool star \citep{Goulding2012,Rackham2018,Rackham2019}, and groups of spots can be quite long lived, with coherence lasting from 10s to 100s of days or more, often over many stellar rotations \citep{Davenport2020,Robertson2020}.

Spots, faculae, and other inhomogeneities on the stellar surface make the interpretation of transmission spectroscopy data more challenging. Transmission spectroscopy is inherently a differential measurement -- for each wavelength bin, the transit depth is measured from the difference between the brightness of the star in transit compared with out of transit. Spots and faculae can cause the stellar spectrum to provide an imperfect representation of the source of light occulted by the planets, primarily because the spectrum of the stellar cord that the planet transits differs from the average stellar spectrum \citep{Rackham2018}. This is further complicated by the fact that the star is rotating and the stellar photosphere visible to an observer changes on the timescale of a transit. If stellar inhomogenities are not correctly accounted for, residuals from an imperfectly subtracted stellar spectrum can be left in the observed transmission spectrum. Under or over corrected stellar features can  confuse the interpretation of absorption features in exoplanet spectra.

Not correcting for this contamination from the star can lead to under or oversubtracted residuals to appear in the transmission spectrum.

The problem of residuals in the stellar spectrum contaminating a transmission spectrum has been identified and studied previously by several groups \citep{Pont2008,Bean2010,Sing2011,Aigrain2012,Huitson2013,Jordan2013,Kreidberg2014,McCullough2014,Barstow2015,Nikolov2015,Herrero2016,Zellem2017,Rackham2018,Rackham2019}. While, there are challenges in the interpretation of transmission spectra of FGK stars \citep{Rackham2019}, M-dwarfs are particularly problematic due to the presence of molecular water in their atmospheres \citep{Jones1995,Rojas2012}. The detection of water in exoplanet atmospheres is of particular interest, both because water is important in atmospheric physics, but also because it is often the strongest absorber at wavelengths we can probe with space telescopes \citep{MillerRicci2009} and can therefore be used to infer the atmospheric mean-molecular weight and scale height.
Stellar contamination of a transmission spectrum can inject a signal of water absorption where none is present, artificially enhance the depth of water absorption features, or even mask the presence of an atmosphere in the data.

NASA's K2 mission \citep{Howell2014} observed the M3V star K2-18 (EPIC 201912552, WISE J113014.45+073516.8, 2MASS J11301450+0735180) in Campaign 1 as a result of being proposed by eight Guest Observer programs\footnote{The eight Guest Observer programs that proposed K2-18 were GO1006 (PI A. Scholtz), GO1036 (PI I. Crossfield), GO1050 (PI A. Sozzetti), GO1051 (PI B. Rojas-Ayala), GO1052 (PI P. Robertson), GO1059 (PI D. Stello), GO1063 (PI V. Sanchez Bejar), and GO1075 (PI B.-O. Demory).}. As a relatively bright (Kp=12.5), cool dwarf star (Teff=3500 K), K2-18 was an excellent target to search for planets. 

\begin{figure}
    \centering
    \includegraphics[width=0.48\textwidth]{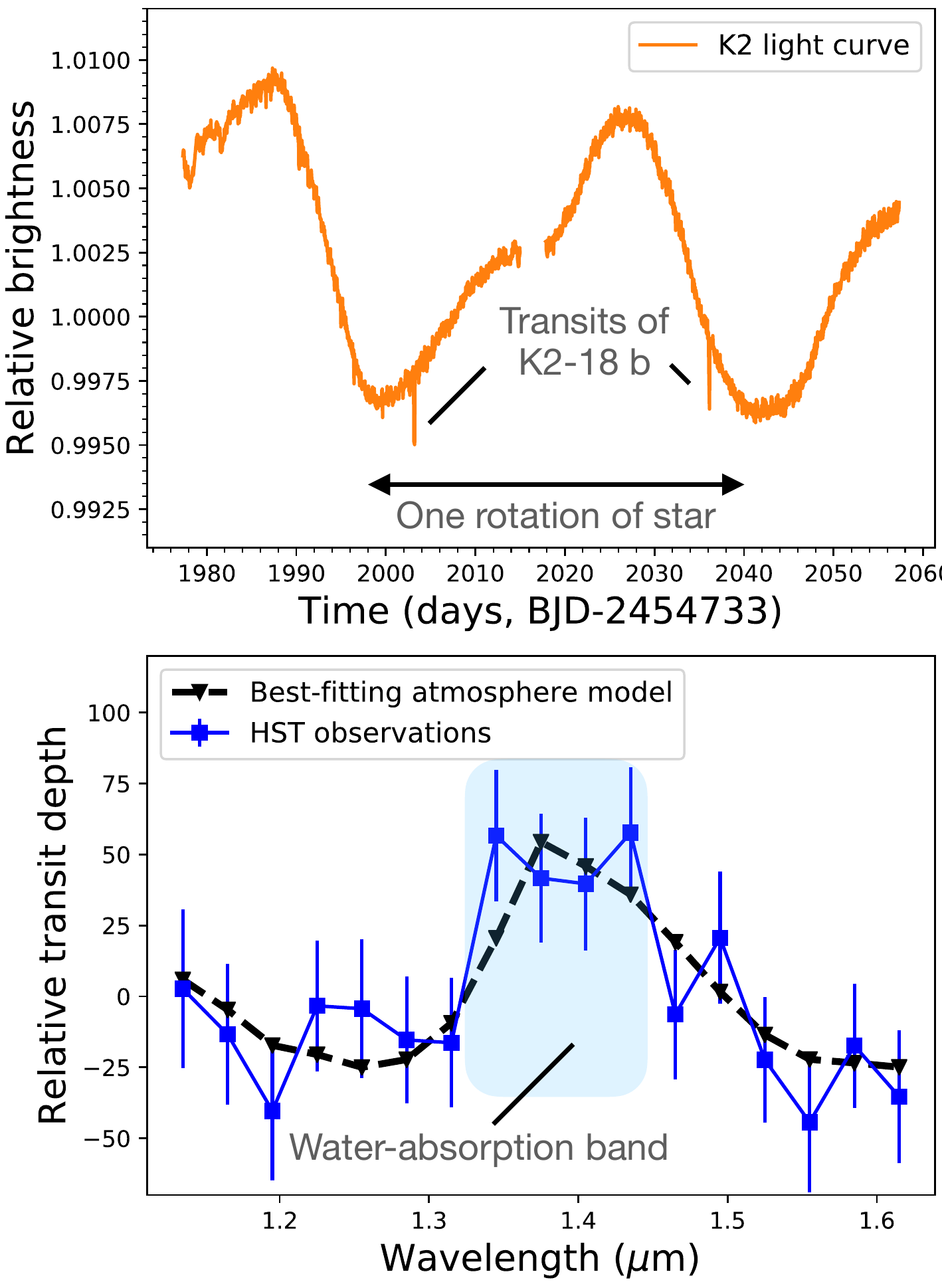}
    \caption{Observed data from K2-18 show a planet with an orbital period of 33 days, a stellar rotation period around 40 days, and an absorption feature in the transmission spectrum. The upper panel shows the K2 mission light curve of K2-18 with the modulation in brightness owing to spots on the stellar surface. The lower panel shows the HST transmission spectrum of K2-18 b measured by \citet{Benneke2019} along with their best-fitting exoplanet atmosphere model. The water absorption band is marked.}
    \label{fig:previous}
\end{figure}

K2-18~b was first reported as a candidate planet by \citet{ForemanMackey2015}. \citet{Montet2015} later confirmed K2-18~b as a sub-Neptune-sized planet with a 33 day orbital period orbiting within the circumstellar habitable zone around K2-18. 
A stellar rotation period of approximately 40 days \citep[as shown in Figure~\ref{fig:previous} and previously reported by ][]{Cloutier2017} can be seen clearly in the K2 data with an amplitude of approximately 1\% indicating that K2-18 has at least a moderate level of magnetic activity and may well be more active than most stars with a similar rotation period \citep{Yang2017}. The light curve of K2-18 is shown in the upper panel of Figure~\ref{fig:previous} demonstrating the orbital period and revealing the two transits. The light curve shown is from the K2SFF high-level science product \citep{Vanderburg2014} and was downloaded from the MAST archive. 

Additional transits of this planet were collected by Spitzer, enabling a refined ephemeris to be measured \citep{Benneke2017}. A mass of $8.6\pm1.4$ M$_\oplus$ was determined using the HARPS and CARMENES spectrographs \citep{Cloutier2017,Sarkis2018}, along with the detection of a second planet orbiting interior to K2-18~b. With these additional data, along with a revised distance estimated from a Gaia parallax, the radius of K2-18~b was revised to $2.71\pm0.07$ R$_\oplus$.

The relatively low density of the planet ($2.4\pm0.4$ g/cc), infra-red brightness of the host star (H=9.1), and deep transits (2900 ppm) make K2-18~b an excellent candidate for atmospheric characterization. Recently, two teams used data from HST to detect water absorption in the atmosphere of K2-18~b via transmission spectroscopy \citep{Benneke2019,Tsiaras2019}. The \citeauthor{Benneke2019} observations and best-fitting model are shown in the lower panel of Figure~\ref{fig:previous}, and the water absorption band at 1.4 $\mu$m is marked. This discovery marks the first detection of water vapor in the atmosphere of a sub-Neptune-sized planet. However, this result sparked our interest because (i) the spectral type of K2-18 is particularly susceptible to contamination of the transmission spectrum by stellar photosphere inhomogeneities, (ii) the water absorption signal is relatively low amplitude (approximately 80 ppm), and (iii) several HST visits were combined together to make the detection. Both \citet{Benneke2019} and \citet{Tsiaras2019} do study whether the atmospheric detection could result from a contaminated transmission spectrum and find it unlikely. However, it is our view that additional analysis may be necessary because both teams use 1-D scaling relationships and mean spot-coverage estimates from \citet{Rackham2018} rather than forward models of the stellar surface. A deeper investigation into contamination would include changes in the distribution of spots between transits and the impact of the rotation of the star during transit observations.



In this work we create plausible, random stellar spot models that emulate the observed photometric spot modulation seen in K2-18 and explore whether the water absorption signal could be a false positive arising from the inhomogeneities on the surface of the star.

\section{Stellar contamination model}
We developed a software model to simulate and study the effects of stellar contamination on transmission spectra of transiting exoplanets, building upon the methods described by \citet{Kostov2013} and \citet{Rackham2018}. Specifically, given a stellar spectrum, a spot spectrum, a spot coverage fraction, and a spot size distribution, the software creates synthetic spectra as a function of stellar rotation phase. Then, by measuring changes in that spectrum as a function of time, we can assess whether changes in an inferred transmission spectrum under the assumption of a spot-free star could be caused by a planet transiting a spotted star.

The simulations proceed as follows. First, we distribute spots on a stellar surface randomly, drawing based off fixed parameters for spot size distribution and coverage fraction that are specified as input parameters to the model. The values of the spot size distribution and coverage fraction can be set based on observed light curves but are typically not well known for any given star. Next, we rotate the sphere at the rotation rate of the star. At specified rotational phases we can evaluate the model by creating synthetic spectra that are a linear sum of the stellar photosphere spectrum and the spot spectrum, with the coefficients of the two models summing to unity and calculated from the visible stellar surface spot-to-photosphere fraction. We assume that the star rotates as a solid body and the spots are persistent.

\begin{figure}
    \centering
    \includegraphics[width=0.48\textwidth]{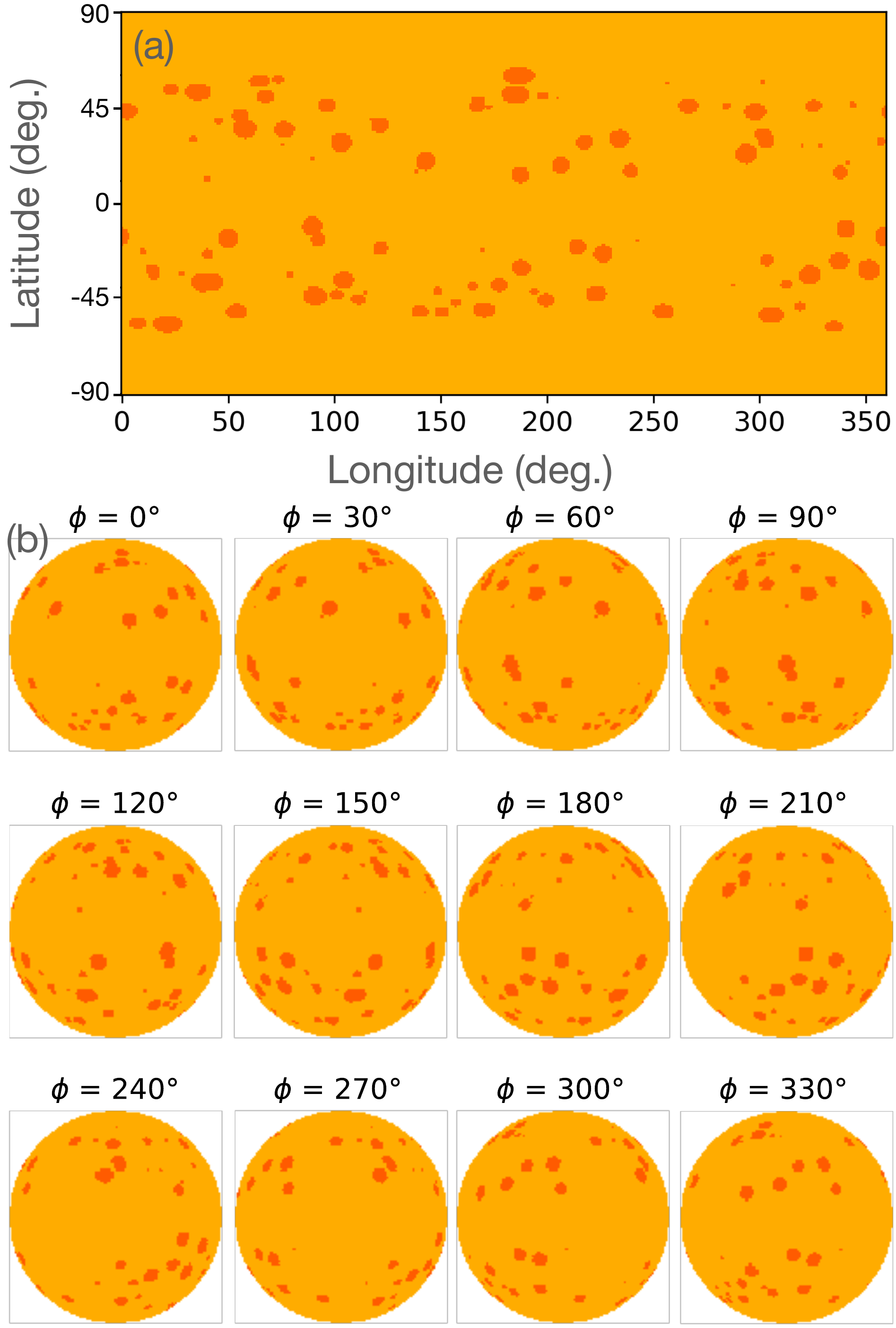}
    \caption{Upper panel: Randomly-simulated equirectangular (latitude-longitude) map with 100 spots covering 5\% of the stellar photosphere. Colors are for illustrative purposes only. Lower panels: The corresponding visible stellar photosphere, orthographically projected and shown as a function of rotation phase.}
    \label{fig:random_map}
\end{figure}

Figure \ref{fig:random_map} shows an example with a randomly-generated equirectangular (latitude-longitude) photospheric map with 100 spots covering $5\%$ of the total surface. The colors are for illustrative purposes only. The map is next projected onto a sphere using orthographic projection, and rotated according to the rotation period of the star. Snapshots of the rotationally-modulated visible stellar photosphere (Figure \ref{fig:random_map}, lower panels) show the spots coming into and out-of-view as a function of the rotation phase, which can in turn be combined with a specific stellar atmosphere model to create synthetic time-series spectra and light curves.

Our model provides a number of enhancements over prior work that relied on first creating a 2-D equirectangular map of the stellar surface with randomly-distributed circular spots for a given spot-covering fraction, and then projecting the map onto a sphere using orthographic projection \citep{Kostov2013,Rackham2018}. This approach incorrectly projects the spots from the input map onto the visible stellar surface (circular spots on the star do not correspond to circular spots on an equirectangular lat-lon map), which is a particularly problematic issue when high-latitude features move into and out-of-view due to rotation. Instead, we account for the projection effect of spots at high latitudes by creating a 2-D equirectangular map using Tissot's indicatrices of deformation, which correctly describe the projection of circles from a sphere to a latitude-longitude map \citep{Tissot1881,Goldberg2007}. The projection is such that circles on a sphere produce distorted ellipses on said map (see upper panel, Figure \ref{fig:random_map}). The primary benefit of this method is that it avoids having higher latitude spots become distorted, avoiding the problem that is seen in some map projections where Greenland appears to be the same size as Africa. Preserving the circular spot-shape is less important because we do not know the specific shapes of the spots we observe -- although most Sun-spots are approximately circular in size. By avoiding distorting the spot shapes, we avoid them having higher weight in the model than is warranted.

In addition to the visible stellar photosphere, we can add to the model an atmosphere-free planet that transits the star. This planet is generally treated as a zero-Kelvin disk that obscures anything that is projected to be behind it, from an observer's perspective.

The model is flexible and enables a large degree of user-defined parameterization, e.g.~utilizing different stellar atmosphere models, instrument wavelength regime and resolution, stellar, planetary and orbital parameters, and changes in the spot patterns. We developed the model to examine whether reported detections of exoplanet atmospheres can be explained by stellar contamination, explore what observations would be required to distinguish between a planetary signal and stellar contamination, and help design strategies for target prioritization based on instrument configuration, stellar type, and wavelength regime. The model is not deterministic, with the spot pattern being randomly drawn. In principle, the code can generate models with faculae and bright spots as well but we did include these in this analysis (although we do discuss this in Section~\ref{sec:brightspots}). We also do not include any limb darkening because it is non-trivial to add a spectral dependence to the multi-component model. 

\section{Building a spot model for K2-18}

Both the K2 data and follow-up observations of K2-18 have provided us with a plethora of information that we can include in our model. We initially built a model to reproduce the K2 light curve of K2-18~by adjusting the spot coverage, spot temperature and spot-size distribution while fixing the stellar spectrum to 3500 K, rotation rate to 38.6 days, and the distance to 38 pc. 



\begin{figure}
    \centering
    \includegraphics[width=0.47\textwidth]{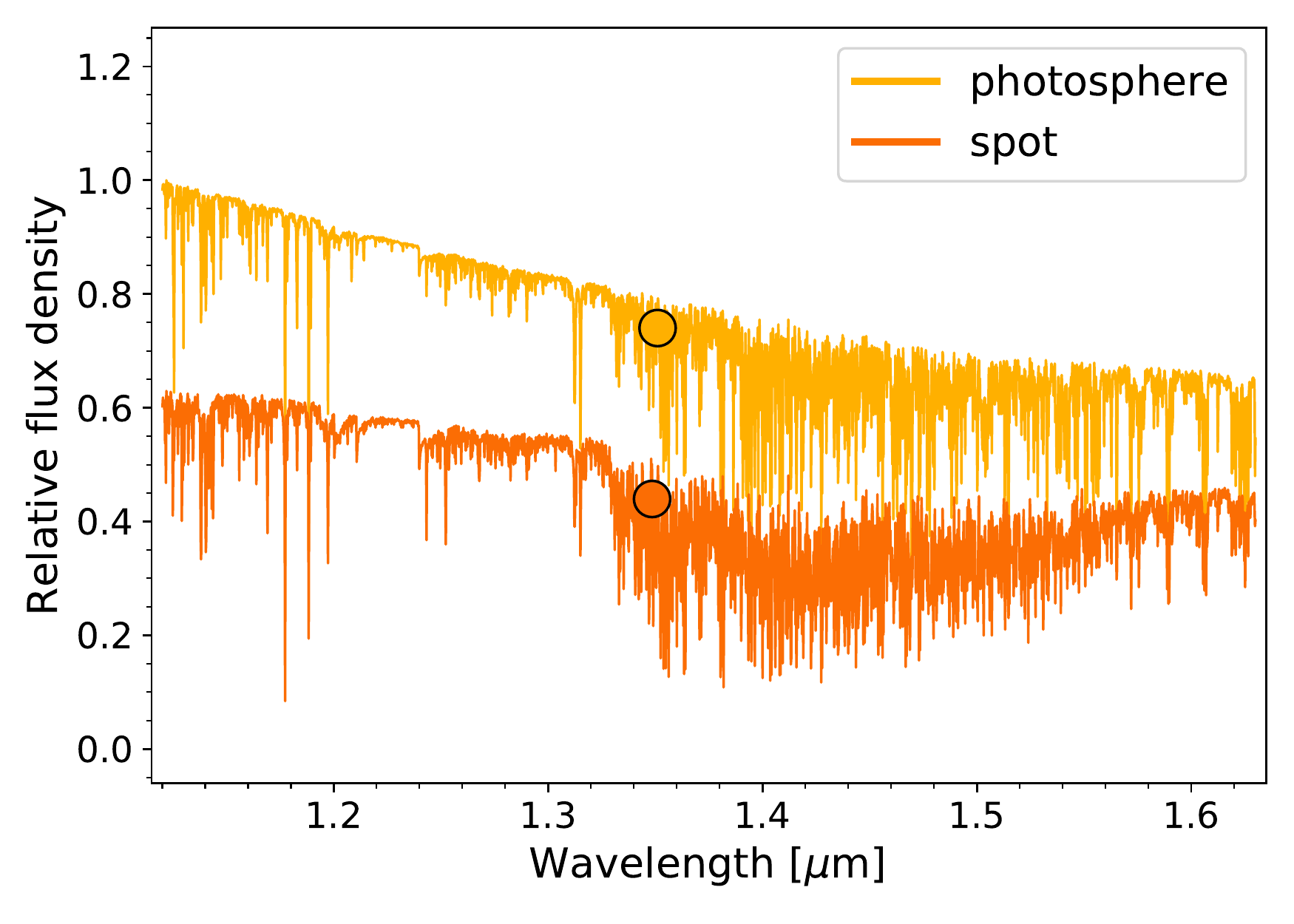}
    \caption{Composite star+spot spectra are produced by combing NexGen stellar atmosphere models. This example shows ${\rm T_{eff} = 3500K, T_{spots} = 3000K}$ in the HST/WFC3 G141 bandpass. These two spectra are summed at every rotation phase with weights according to the visible coverage. The circles are the flux-weighted mean of the bandpass.}
    \label{fig:Fig_setup}
\end{figure}

The spot coverage fraction of K2-18 is not well constrained and likely varies over time, likewise with the spot temperatures and the size distribution of spots. These parameters are degenerate, and also yield different results depending on the precise properties of a given random draw of spot placement. Nevertheless, we can use the photometric variability of K2-18, measured as ${\rm \sim 7-10\ mmag}$ semi-amplitude in the Kepler, $B$ and $R$-bands \citep{Cloutier2017,Sarkis2018}, as the parameter to match with our model. The time between the three photometric data sets span 3 years and we do not see significant changes in the spot amplitude during that time. While these measurements were not taken simultaneously with the HST observations (and all-sky instruments like ASAS-SN lack the precision to see variability in the K2-18 light curve), we assume that the variability of K2-18 has not changed significantly because the photometric data bookend the HST data that was collected in 2015--2017, and are relatively consistent over that timeframe. We experimented by using  NextGen stellar atmosphere models \citep{Hauschildt1999} with a photosphere effective temperature of 3500 K and varied the spot temperatures and coverage fraction. We randomly selected spot sizes from between the largest spots typically seen on the Sun \citep{Mandal2017,Johnson2021} and the largest spots seen on M-dwarfs \citep{Berdyugina2011} (spots that vary between 0.01--0.05 solar-radii). This leads to 100 total spots on the star (see Figure~\ref{fig:random_map}), and that is consistent with the limited observations of M-dwarf spots. We note that using smaller spots requires significantly high spot coverage fraction to replicate the observed variability \citep{Johnson2021}, while larger spots would require coverage fraction of 1--3\% (discussed further in Section~\ref{sec:largespots}). The latitude range of spots is a runable parameter in the model. We restricted spots to being at latitudes with $\pm60^\circ$ of the stellar equator to avoid polar spots. While polar spots may be present on cool stars, the projected size of these spots is small and therefore unlikely to contribute significantly to the total stellar brightness. Through varying the coverage fraction and spot temperature, and integrating over the K2 bandpass, we found that a spot-covering fraction of 5\% and a spot temperature of 3000 K produces typically in the range ${\rm \sim0.5-1\%}$ semi-amplitudes which is in line with observations ($>$70\% of samples had amplitudes in this range). The photosphere and spot spectra are shown in Figure \ref{fig:Fig_setup} for the range of the HST/WFC3 G141 grism. Having a star with spots that are 500 K cooler than the stellar photosphere and cover 5\% of the stellar surface is reasonable for a cool dwarf such as K2-18 \citep{Berdyugina2005,Herbst2021}. The precise observed semi-amplitude in the variability can be as low as a few 10s of parts-per-million and as high as a few percent, depending on the rotational asymmetry of the spot pattern but has an average right in the range observed for K2-18. 

\begin{figure}
    \centering
    \includegraphics[width=0.47\textwidth]{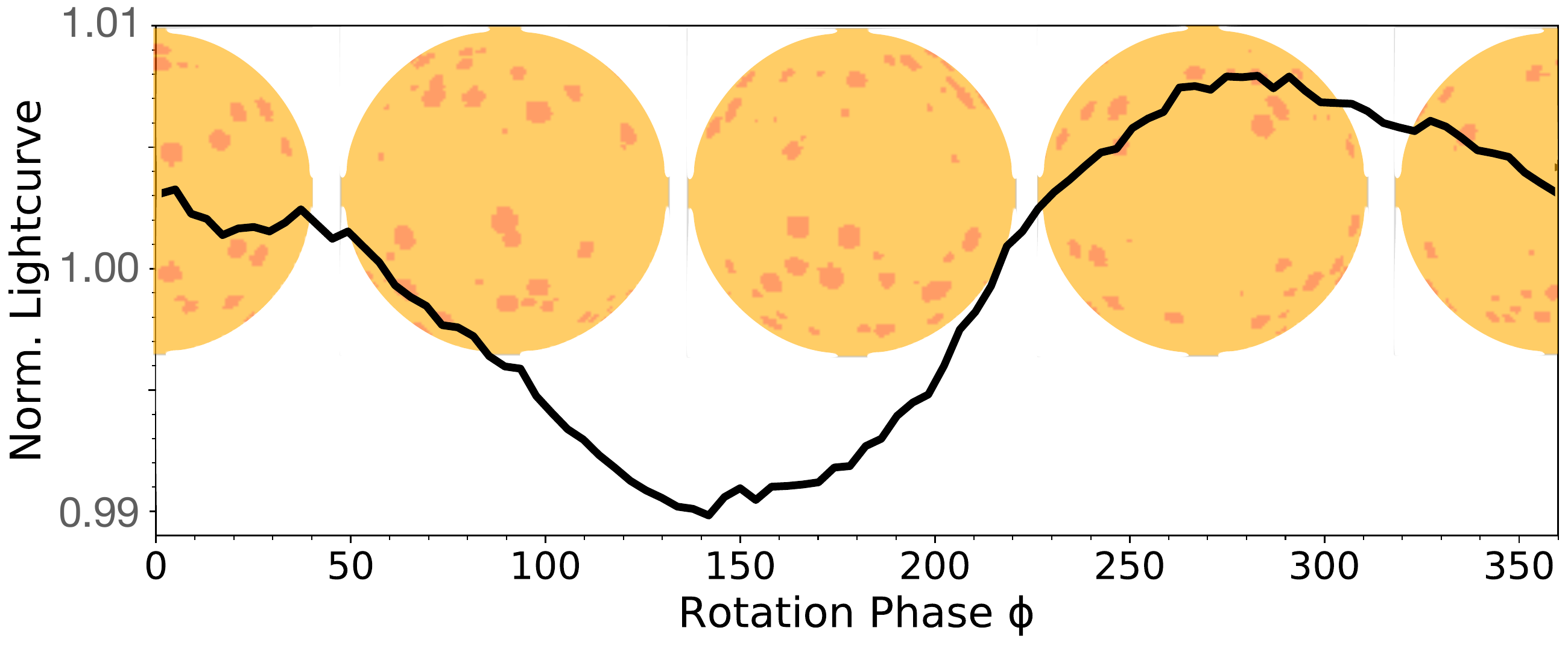}
    \caption{Rotationally-modulated, disk-integrated light curve produced by a variable projected spot coverage fraction. The background images are snapshots of the visible stellar disk at different rotation angles, separated by 90 degrees. The light curve generated by this spot coverage fraction is produced by integrating the combined spot+photosphere spectrum over the Kepler-bandpass. We reproduce the variability amplitude that was measured previously \citep{Montet2015,Sarkis2018} for K2-18 where the rotation period is approximately 40 days. We use these variable spot coverage fractions to create time-variable synthetic stellar spectra.}
    \label{fig:Fig_LC}
\end{figure}

A disk-integrated, rotationally-modulated, white-light light curve is shown in Figure \ref{fig:Fig_LC}, sampled every 10 degrees of stellar rotation. This figure is generated from the same spot pattern as in Figure~\ref{fig:random_map}. We note that extracting a `map' of the visible stellar disk based on photometric variability alone is highly degenerate \citep[e.g.][]{Russell1906}. With that said, the aim of this paper is not to provide a conclusive solution but to explore the plausibility of whether the transmission spectrum of K2-18~b could be contaminated with stellar variability. Our model accounts for two related but separate effects that cause contaminated exoplanet transmission spectra: (a) The planet crossing a chord of the star that may have different spot properties to the average spectrum of the unocculted star, and (b) The star itself is rotating, continuously changing the average spectrum of the star that is being occulted.

\subsection{A simple model of a single transit}

Currently underpinning transmission spectroscopy measurements is the approximation that the host star photosphere is well-understood, and that the disk-integrated stellar spectrum can be inferred before, during, and after a transit, i.e. on timescales corresponding to many degrees of stellar rotation. However, the stellar spectrum is a composite of the photosphere, spots, and faculae, where the latter two vary both spatially and temporally as the star rotates. These brightness variations contaminate the observed spectrum and can mimic or mask atmospheric features interpreted as planetary, for example enhancing the depth of a water feature, masking the presence of water in the data, or injecting a signal of water absorption where none is present. Recent work shows that in some cases stellar contamination can (and does) produce signals up to an order of magnitude larger than the intrinsic planet features \citep{Rackham2018,Zhang2018}.

\begin{figure}
    \centering
    \includegraphics[width=0.5\textwidth]{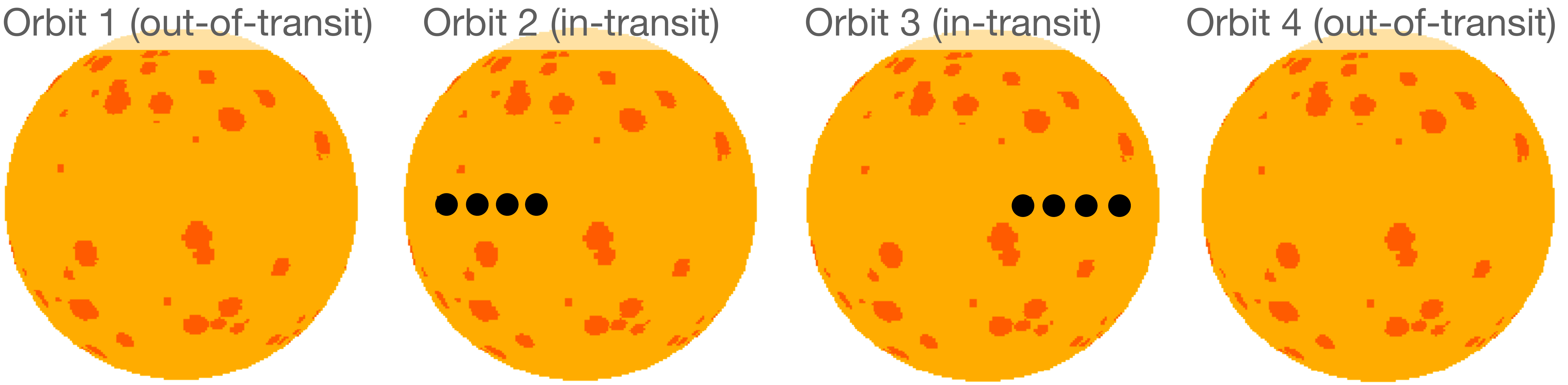}
    \caption{Visible stellar photosphere for a single simulated HST transit observation with 4-orbits, and 4-exposures-per-orbit. A K2-18-like planet transits the star, for this illustration we placed four exposures during a given orbit on the same disk image. The stellar rotation phase was randomly selected using the photosphere map of Figure~\ref{fig:random_map}. The black dot represents a transiting planet with no atmosphere, moving across the stellar disk; the star is rotating during the observations but only turns by a little over 1$^\circ$. The stellar and the planetary disks are to scale, and represent a 2.7 ${R_\oplus}$ planet on a 33-day orbit transiting a 0.4 ${R_\odot}$ star. For simplicity, we used a zero inclination transit.}
    \label{fig:visit1}
\end{figure}

During transit, a planet moves across the disk of the host star and blocks only a fraction of the stellar surface. The surface blocked may not be representative of the star as a whole, this is known as the transit light source effect \citep[e.g.][]{Rackham2018}. Simultaneously, the star itself rotates and different features move in and out of view. This is demonstrated in Figure~\ref{fig:visit1} which shows a simulated 4-orbit, 4-exposures-per-orbit HST visit (representing a single transit) of a K2-18-like system during a transit occurring at a randomly-selected rotation phase (using the same photosphere map as shown in Figure~\ref{fig:Fig_setup}). Each image shows the rotating stellar disk over the course of one HST orbit. While the stellar rotation is not noticeable on the scale of the image for this particular simulated system (the rotation period of ${\sim925}$ hours is 2+ orders of magnitude longer than the transit duration of ${\sim3.5}$ hours), it is nevertheless important for the emerging stellar spectrum as discussed below. In general, the closer the rotation period is to the duration of the transit, the more pronounced the impact. For the particular configuration shown in the figure, the planet transits during orbits 2 and 3, without occulting any spots. However, in general, spot-crossing events can occur and have a strong effect on both the measured light curve and the spectrum. By design, our model naturally accounts for this effect. 

\subsection{A model for the 8-transit HST observations}

\begin{figure}
    \centering
    \includegraphics[width=0.48\textwidth]{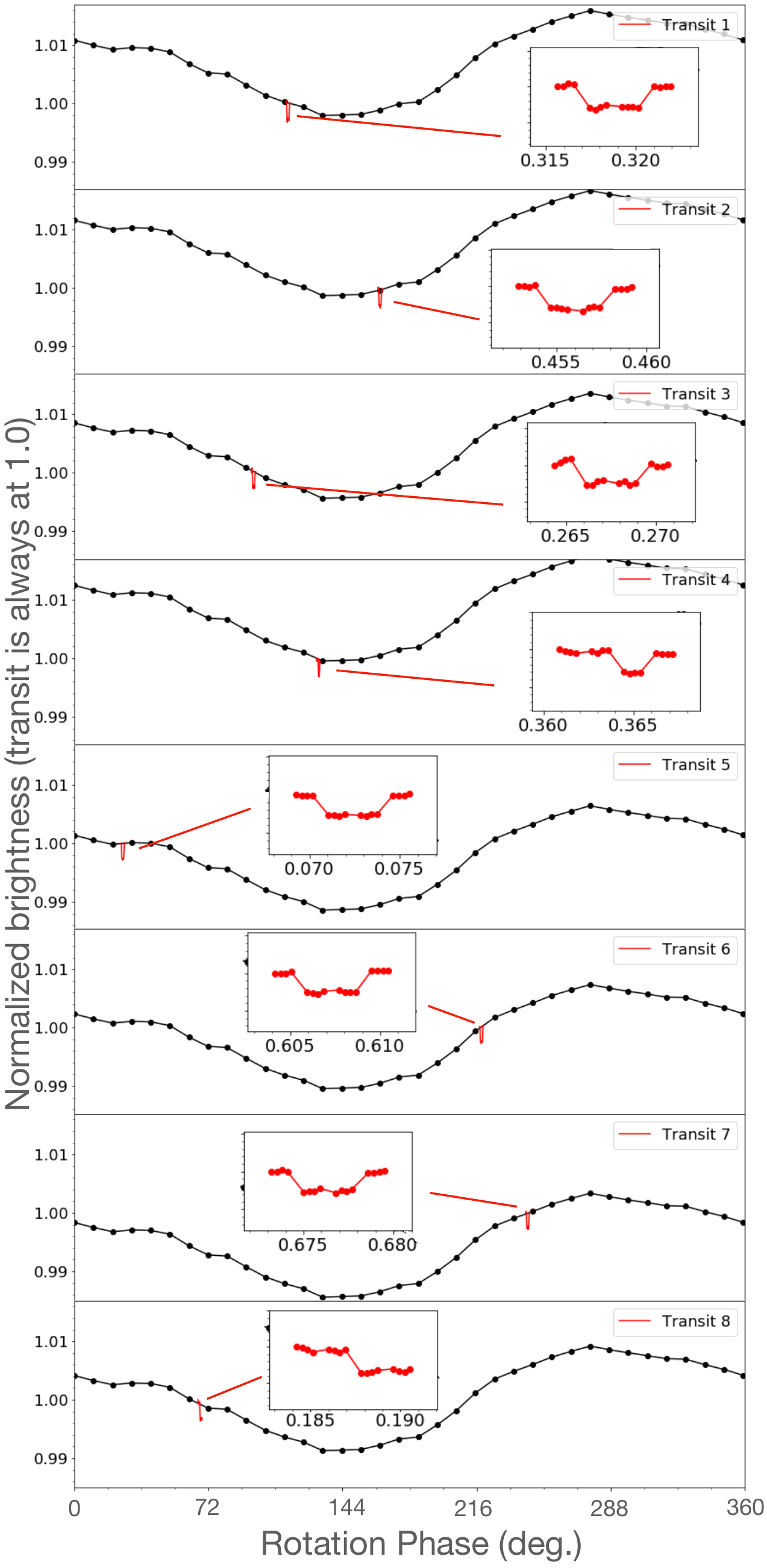}
    \caption{Stellar light curves (black symbols) as a function of stellar rotation phase for 8 planet transits of the same star, simulating the observations from HST of K2-18~b. The rotation light curve in this example is fixed to be the same as in Figure~\ref{fig:Fig_LC}, starting with the one shown in Figure~\ref{fig:visit1}, and following the observing strategy of K2-18. The red symbols (also inset panels) represent 8 simulated HST transit observations at 8 randomly-selected rotation phases. }
    \label{fig:Fig_LC_8visits}
\end{figure}

The HST observations of K2-18~b cover eight transits collected over two years \citep{Benneke2019}. To simulate the HST observations, we created eight synthetic transit observations with eight randomly-selected rotation phases, while accounting for the observing strategy used. K2-18 rotates through one degree during the transit of K2-18~b, and a total of approximately 2 degrees over each HST observation. Specifically, transits 1, 2, 3, 5, 6, and 7 cover OOT-ITR-ITR-OOT, transit 4 covers OOT-OOT-ITR-OOT, and transit 8 covers OOT-OOT-ITR-ITR, where OOT are Out-Of-Transit and ITR are In-Transit sections of the light curve. This is shown in Figure~\ref{fig:Fig_LC_8visits}, where the inset transits correspond to said eight realizations (the top panel represents the configuration of the system shown in Figure~\ref{fig:visit1}). For the demonstration presented here, we assume that the spot pattern does not change between the different transits. In general, our software can include a changing spot pattern between different transits. However, using a random or fixed spot pattern does not change any results because each projected face of the star is essentially random.
The synthetic Kepler-bandpass spot patterns corresponding to the transits shown in Figure~\ref{fig:Fig_LC_8visits} cover the full-rotation of the star. For example, transit \#1 is at a rotation phase of 295$^\circ$ and transit \#2 is at a rotation phase 162$^\circ$. 

\begin{figure}
    \centering
    \includegraphics[width=0.48\textwidth]{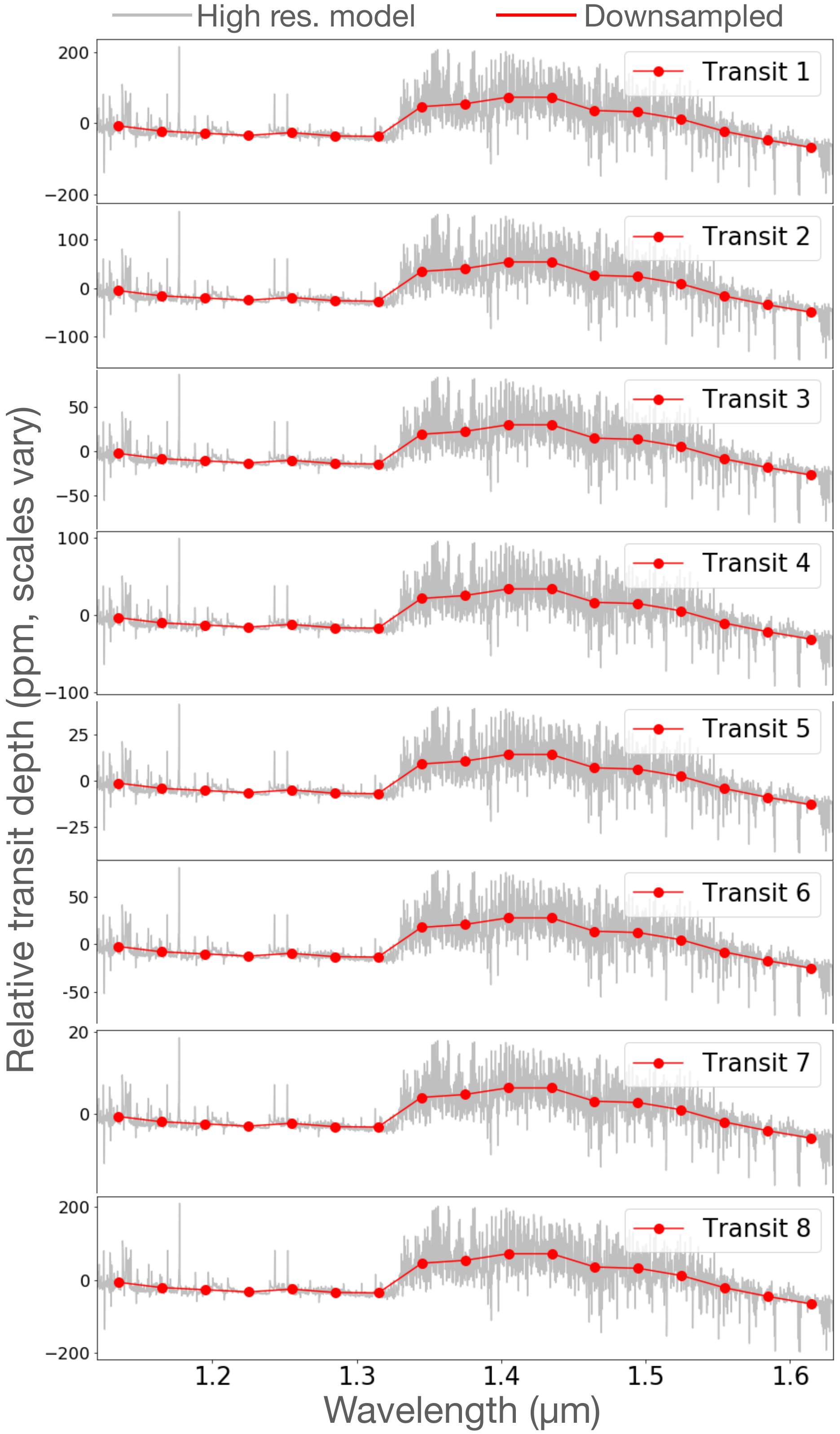}
    \caption{Synthetic contaminated transmission spectrum (gray, full resolution; red, downsampled to the  HST/WFC3 G141 filter), in terms of (OOT-ITR)/ITR, mean-subtracted and on the scale of ppm, for the 8 transits outlined in Fig \ref{fig:Fig_LC_8visits}. Note the y-axis changes between transits. In half the transits the stellar contamination signal in the transmission spectra is low ($<$50 ppm) but in several of the transits the contamination signal could dwarf any atmospheric signal.}
    \label{fig:AvgSpec}
\end{figure}

Next, assuming a transiting planet with no atmosphere, for each simulated transit of each realization we calculate the rotationally-modulated synthetic contaminated transmission spectrum in terms of (OOT spectrum - ITR spectrum)/ITR spectrum, accounting for the loss of light due to the fraction of the stellar disk covered by the transiting planet. The latter effect naturally produces a ${\sim 2900}$ ppm depth transit (see Figure \ref{fig:Fig_LC_8visits}), consistent with the measured average transit depth of K2-18~b in the HST/WFC3 G141 filter \citep{Benneke2019,Tsiaras2019}. The simulated contaminated transmission spectrum for a set of eight transits is shown in Figure~\ref{fig:AvgSpec}, showing variability in the difference spectrum from one transit to the next at the $\sim50$ ppm level. The most prominent variability occurs between 1.3--1.4 $\mu$m which is precisely the region where atmospheric absorption by water molecules is most prevalent. Each synthetic transmission spectrum is calculated at high spectral resolution, then downsampled to the resolution presented by \citet{Benneke2019}. The downsampling is performed by summing over each wavelength range. There is a reasonable amount of variety in the level of variability between the synthetic spectra for each transit. However, this variability would not easily be identified in an individual HST observation because typical uncertainties per spectral bin, per transit seen were 60--70 ppm. Only by combining multiple observed transit spectra would the signal be significantly detected. 

\begin{figure}
    \centering

    \includegraphics[width=0.48\textwidth]{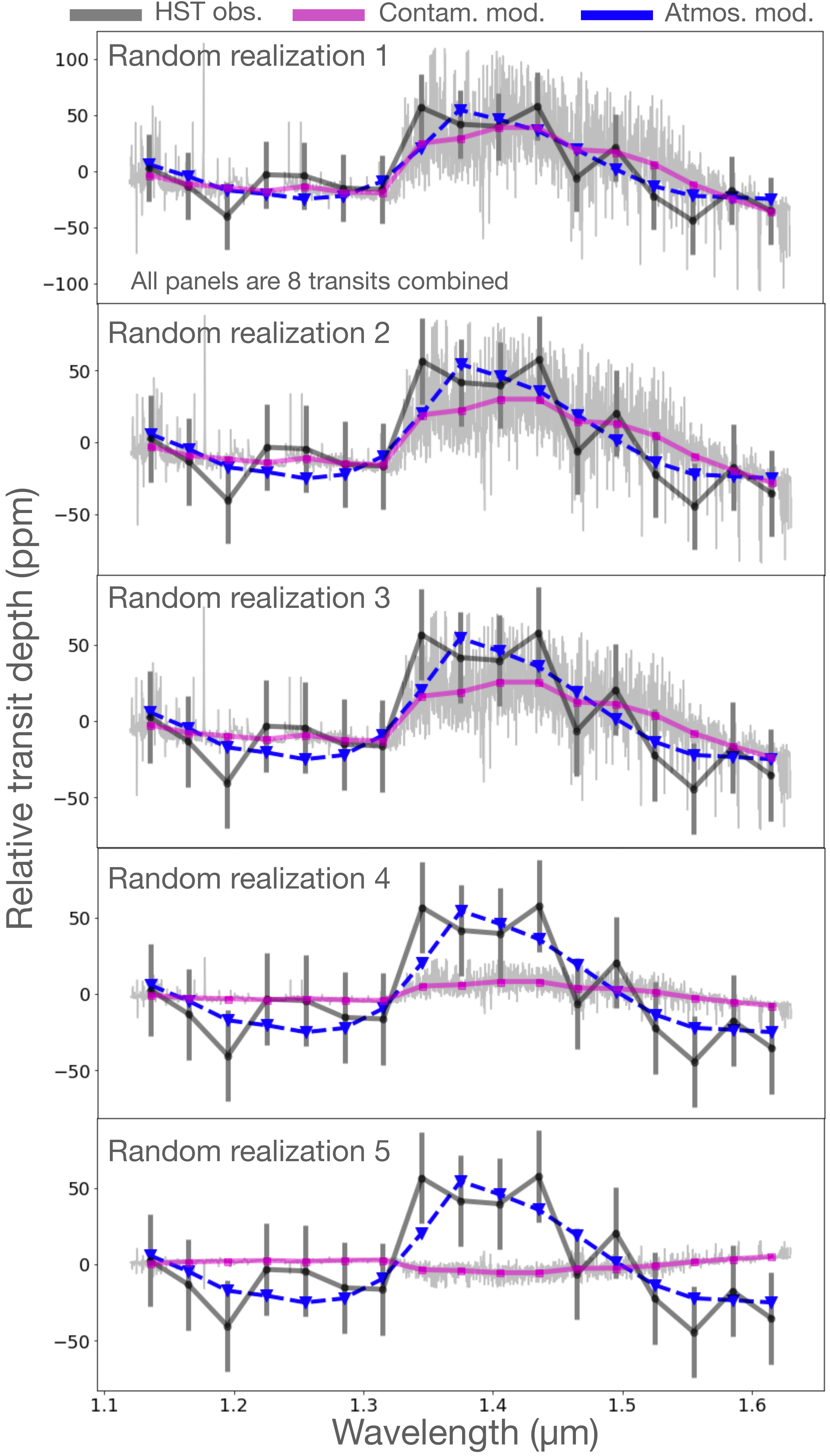}
    \caption{Five random transmission spectra generated by our spot model code demonstrate that some models show good agreement with the observations. Each panel shows one realization of an combined 8-transit spectrum. The purple data are the spot contamination model, the HST observations for K2-18 from \citet{Benneke2019} is shown in grey, and the best-fitting exoplanet transmission spectrum is shown in blue. The top panel represents the combination of the individual transits shown in Figure~\ref{fig:Fig_LC_8visits} while the remaining spectra are different random simulations. Some of the simulated spectra are quite different from the observations (e.g. panels 4 and 5 here). However, many of the models produce a prominent water feature---created by stellar contamination alone---as shown in the top 4 panels.}
    \label{fig:Fig_avg_spec_all_visit_10_iter}
\end{figure}

For each transit of each realization we sum the simulated contaminated transmission spectrum to the HST/WFC3 G141 spectral resolution and create an 8-visit-average difference spectrum for comparison with the observations. Figure~\ref{fig:Fig_avg_spec_all_visit_10_iter} shows five combined 8-transit transmission spectra, where the purple line represents the simulated difference spectrum and the gray line represents the HST/WFC3 G141 observations of K2-18~b. The top panel of Figure~\ref{fig:Fig_avg_spec_all_visit_10_iter} represents the results from the simulation based on the map from Figure \ref{fig:random_map} and the eight transits from Figure \ref{fig:Fig_LC_8visits}, the remaining panels represent the results from four other iterations of the model drawn at random (based on their respective random stellar maps and eight rotation phases). 

Iterations of the model -- but with the same input parameters (i.e. number of spots, spot-covering fraction, photosphere and spot spectra) -- with different, randomly-generated stellar maps and rotation phases during transits, produce different 8-visit-averaged contaminated transmission spectrum. This is clearly seen in Figure \ref{fig:Fig_avg_spec_all_visit_10_iter}, where the model produces a variety of difference spectra depending on the spot distribution and rotation phase at transit; sometimes the difference spectra even exhibit an inverted shallow feature. The simulated difference spectra fairly often show a prominent feature that may mask or mimic the potential detection of water in the atmosphere of a transiting planet -- and thus contaminate the transmission spectrum. In particular, the top three panels in Figure~\ref{fig:Fig_avg_spec_all_visit_10_iter} show a significant feature between 1.3 and 1.4 $\mu$m, and the simulated spectral models are in very good agreement with the HST data. The star is solely responsible for any features in the simulated difference spectrum -- the transiting planet in the model has no atmosphere -- and the simulated difference spectrum in Figure~\ref{fig:Fig_avg_spec_all_visit_10_iter} is not a fit to the observed difference spectrum but random draws from the model. 

\subsection{Assessing how well the HST observations are described by our model}

\begin{figure*}[htb]
    \centering
    \includegraphics[width=0.85\textwidth]{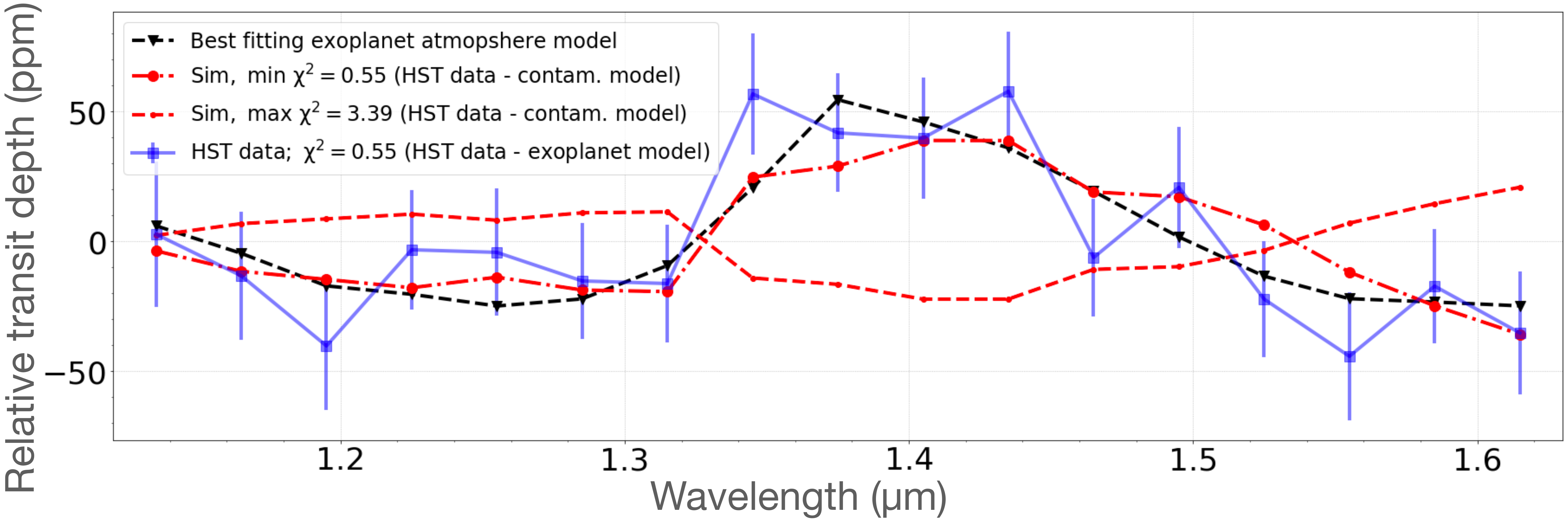}
    \caption{The spot model with the lowest reduced $\chi^2$ is a better fit to the observed data than the best fitting exoplanet model. The best and worst fitting models (compared with the observed data) from a set of 1000 simulations are shown in red (best fitting: ${\rm \chi^2_{min} = 0.55}$, red dot-dashed; and worst fitting ${\rm \chi^2_{min} = 3.39}$, red dashed). For comparison, the reduced ${\rm \chi^2}$ between the \citet{Benneke2019} data (blue) and their best fitting exoplanet model (black dashed) is ${\rm \chi^2 = 0.55}$. Half of the spot model spectra have a reduced $\chi^2 < 1$ and 1\% fit better than the best fitting exoplanet atmosphere model.}
    \label{fig:gof}
\end{figure*}

To study the significance of these similarities, we ran 1000 random draws from the model like those shown in Figure~\ref{fig:Fig_avg_spec_all_visit_10_iter}. For each draw, we calculated the reduced $\chi^2$ between the simulation and the \citet{Benneke2019} observations (with $\chi^2 = \sum (y - m)^2 / \sigma^2$ and reduced $\chi^2 = \chi^2 / \nu$ and $\nu$=17, the number of degrees of freedom). The distributions of the reduced $\chi^2$ for each of the 1000 random draws ranges from 0.5--3.0. Half of the simulated difference spectra, induced by stellar contamination alone, produce reduced $\chi^2$ with respect to the HST data of smaller than one, and 10\% have a reduced $\chi^2$ less than 0.65. Figure~\ref{fig:gof} shows the best fitting and worst fitting draws along with the best-fitting exoplanet atmosphere model from \citet{Benneke2019}, shown in black. Just over 1\% of random draws yield a better fit to the data than the best-fitting exoplanet atmosphere model. This demonstrates that the observed data could be explained by the model we have generated.

\begin{figure}[htb]
    \centering
    \includegraphics[width=0.48\textwidth]{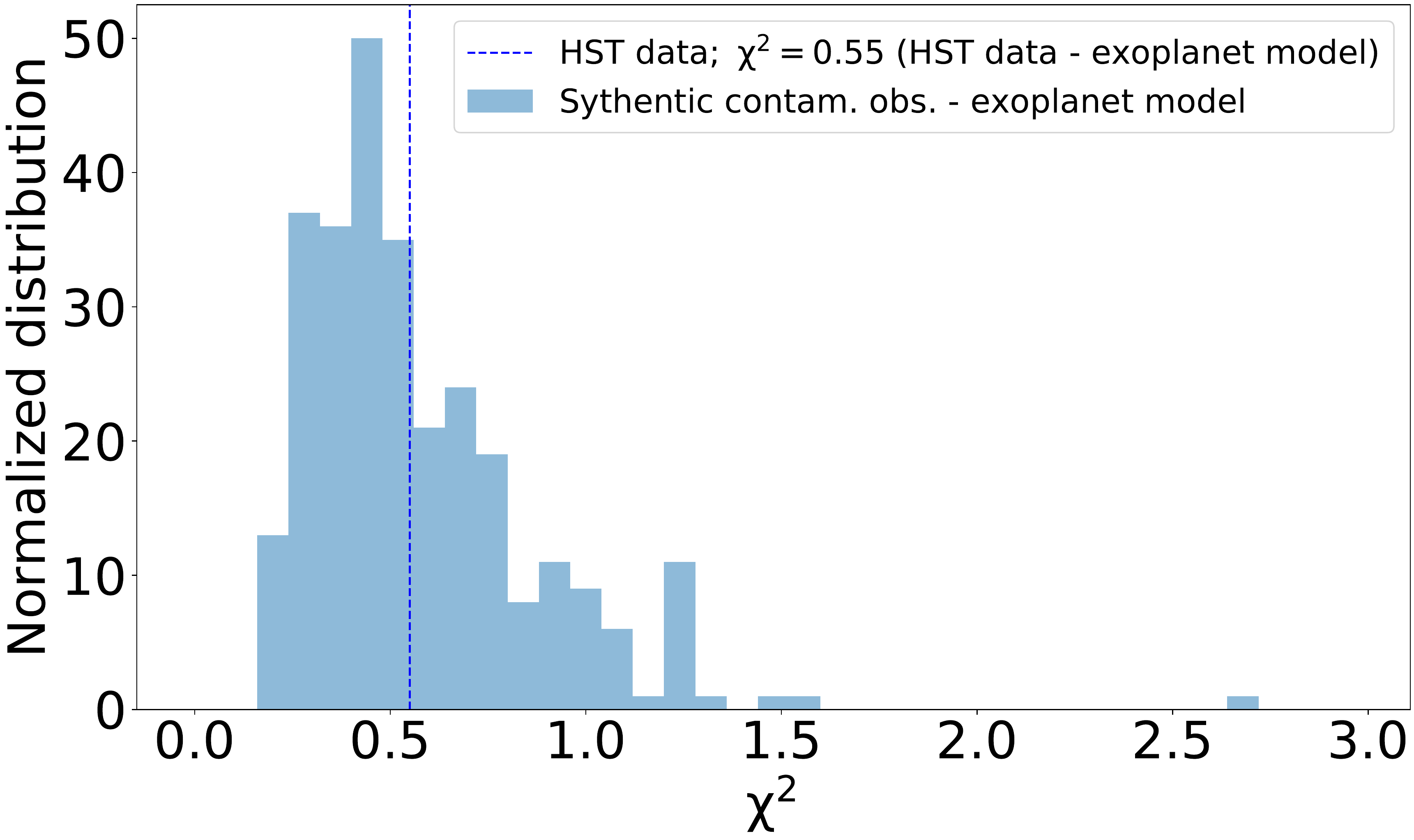}
    \caption{Using the spot model samples to generate synthetic observations shows that a large fraction of observations would provide good fits to an exoplanet atmosphere model. We resampled the spot model spectra to include observational uncertainty and find that 40\% of synthetic observations fit the best-fitting exoplanet atmosphere model from \citet{Benneke2019} at least as well as the true data.}
    \label{fig:hist_}
\end{figure}

We also performed the reverse experiment. We treated the model spectra as synthetic HST observations, resampling them to include an observational uncertainty using the measured uncertainty reported by \citet{Benneke2019}. We then compared the reduced $\chi^2$ of the synthetic observations to the best-fitting exoplanet atmosphere model. Figure ~\ref{fig:hist_} shows the $\chi^2$ of those samples. We found that $>$40\% of simulated observed transmission spectra generate a reduced $\chi^2$ lower than that between the real HST observations and the exoplanet atmosphere model (reduced $\chi^2$=0.55). This demonstrates that under the assumption that our stellar spot model is suitably realistic, and the probability of detecting an atmosphere where none is seen is rather high.

\section{Discussions}

\subsection{Limitations in the spot model}
The model we have developed is certainly far from complete, and therefore we need to assess whether the limitations of the model could be biasing our results or even generating the contamination signal where none should exist. In particular limitations include: the star rotates as a solid body, rather than differentially; there are no faculae or spots of variable temperature; the sizes of the spots is poorly constrained; and the star is not limb darkened. Spots are also fixed in location and do not evolve as is observed on some stars. We explored these effects in more detail to understand the impact of not including them. We do not expect differential rotation to have a large impact on these results because the spots are still rotating into and out of view, just at slightly different rates. Likewise, the impact of limb darkening should be limited because the brightness change from edge to limb is relatively small. We experimented with using random photospheres with every transit but the results did not change. However, the issue of faculae and bright spots, and spot size does warrant closer inspection. 

\subsection{Faculae and bright spots}
\label{sec:brightspots}
Our spot model is capable of generating bright spots, but the primary reason for not including them is that there are not any especially good constraints on their coverage fraction, temperature, and size distribution. \citet{Wakeford2019} found that 5800 K hot spots may be present on a few percent of the surface of TRAPPIST-1, although they also find a spot coverage fraction of 60\% which is far beyond what would be expected for K2-18 and therefore likely not directly comparable to K2-18. On the other side of the mass range, \citet{Montet2017} found that stars in the Kepler data (mostly FGK stars) that rotate with periods longer than 25 days have light curves that are faculae-dominated. To explore this we experimented with including 20 bright spots that cover 0.5\% of the photosphere (and no dark spots). We found that a spot temperature of 4000 K gave us the right amplitude in the light curve to match the photometry data sets. Transits of bright spots are very obvious in the time series of the transits and are frequently seen in the simulated models. Unocculted bright spots occasionally produced significant peaks in the transmission spectrum at 1.3--1.4 $\mu$m but they are much less frequent than for cool spots with lower amplitudes - just 1 in 5 had amplitudes over 20 ppm, and only a few percent produced a peak over 50 ppm. Therefore, a star with only bright spots is unlikely to have produced the signal seen in the K2-18 transmission spectrum, although it cannot be ruled out. 

We next tried including dark and bright spots. With 4.5\% coverage from cool spots and 0.5\% coverage from bright spots we essentially reproduce the results for the dark spot model, with a marginally higher probability of seeing a significant peak around the water absorption band. Unless bright spots are far more prevalent than dark spots on K2-18, we find it unlikely that our spot model is incorrectly generating a contamination signal where none should exist due inaccurate modeling of faculae versus spots.

\subsection{Spot size distribution}
\label{sec:largespots}
We have assumed that the spot-sizes on K2-18 range in size from 0.01--0.05 solar-radii, which is approximately the range between large spots on the Sun to the largest seen on M-dwarfs. However, if spots are much larger than we have simulated, or if they are strongly clustered (which has the same effect as large spots) then our results may not be valid. We examined a case of having 20 large spots, which can lead to spots larger than $30^\circ$ in diameter. An example of a star with large spots in shown in Figure~\ref{fig:spot_crossing_1}. In this case we need lower spot covering to match the amplitude seen in K2 data. We find a coverage of 1--3\% yields K2-band semi-amplitudes of 0.5--1.5\%. The impact here is that the contamination signal seen in the IR transmission spectra is generally much lower, and not generally consistent with the HST observations of K2-18. We simulated 500 large spot models and found that all of them provided a poorer-fit to the HST data than the best-fitting exoplanet atmosphere model from \citet{Benneke2019}. The typical amplitudes seen with the large-spot spectra are approximately half that seen with the standard spot model. Therefore, if spots are generally much larger than expected or are heavily clustered then the exoplanet atmospheric detection in K2-18 would be the heavily favored model. 

\subsection{The impact of model image resolution}
By design, the visible photosphere of our simulated stars is pixelated and thus contains an intrinsic resolution. The simulations presented in this work are 180x180 pixels in size; the models quickly become computationally prohibitive as the pixel resolution increases. To evaluate the impact of this pixelation on the results, we tested additional resolutions: 60x60 pixels, 180x180 pixels, 360x360 pixels, 540x540 pixels, 720x720 pixels, 960x960 pixels, and 1640x1640 pixels (spanning nearly 2 orders of magnitude increase in resolution). There is typically very high variance at 60x60 resolutions and therefore there is noise generated by the resolution. However, while the reduced chi-square between the corresponding synthetic contaminated stellar spectrum and the \citet{Benneke2019} model varies a little with the resolution, the change is small at resolutions of 180x180 and above, and levels off beyond 540x540 pixels. Therefore, we find that our model is run at an appropriate resolution.

\subsection{Spot crossings}
\label{sec:spotcrossing}
\begin{figure*}
    \centering
    \includegraphics[width=0.78\textwidth]{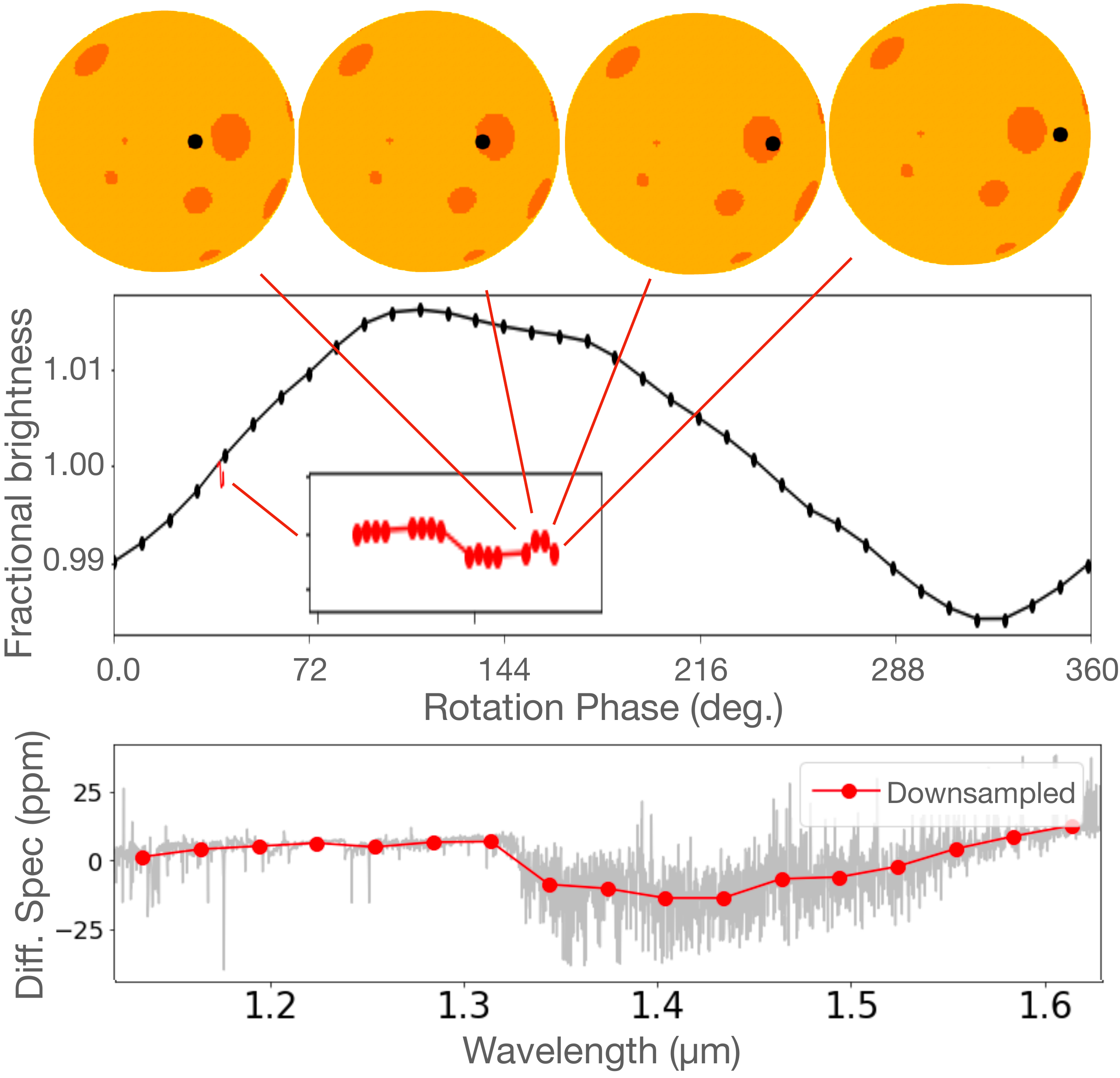}
    \caption{An example spot crossing transit that produces a clear event in the transit light curve and transmission spectrum. In this example, the fourth synthetic HST orbit in the transit observation crosses a spot. The spot crossing is visible as a significant increase in the flux measurement during the spot crossing (central panel). The lower panel shows the resulting difference spectrum ([in-transit - out-of-transit] / in-transit) where the inverse of a typical transmission feature is seen. This shape is characteristic of a spot crossing event. This example is for the large spot case, to show the effect on the transit shape more clearly.}
    \label{fig:spot_crossing_1}
\end{figure*}

In our simulations, spot crossings during transits occasionally occur. These are relatively easy to identify in the integrated light curves and transmission spectra. Such an occurrence is illustrated in Figure~\ref{fig:spot_crossing_1}, where the planet crosses a spot (shown for the large spot model because the spot crossing event is easier to see in these data). The corresponding light curve exhibits the well-known in-transit brightening present in many observed exoplanet transits \citep{Morris2017}. Such events are not detected in any of the HST observations of K2-18~b \citep{Benneke2019,Tsiaras2019} and thus, for the purposes of comparing said observations to the results presented here, we do not think that HST observations have any spot-crossings (or at least no transits of large spots). Synthetic observations that include spot-crossing do not generally provide a good fit to the observed data.
However, it is worth noting that transmission spectra that cross a spot tend to have a particular shape -- they have an inverted spectrum compared with non-spot crossing transits that resembles an inverted water absorption feature. Should the spot crossing not be detected any true water absorption seen in the transmission spectrum would be muted by contamination from the spot.

The lack of any spot crossings detected in the HST data is not especially surprising. With larger spots discussed in Section~\ref{sec:largespots} spot crossing are fairly frequent, but with the smaller spots used in the primary study only 10\% of simulations have a transit cross a spot that produces a detectable spot crossing event (a larger fraction have a crossing of a very small spot but those do not generally produce a measurable effect). If there are many small spots at the transit latitude then we might expect additional noise during the HST transit compared with the out of transit data, but this is not seen - the scatter on the data appears consistent during each visit \citep{Tsiaras2019}.



\subsection{Out-of-transit brightness as a constraint on spot-coverage}
In our analysis to this point we have assumed that the K2-18 HST observations were collected at a random rotational phase of the star. However, we do have some limited constraints; we can measure the out-of-transit brightness of K2-18 that was observed by HST. 

\begin{figure}
    \centering
    \includegraphics[width=0.5\textwidth]{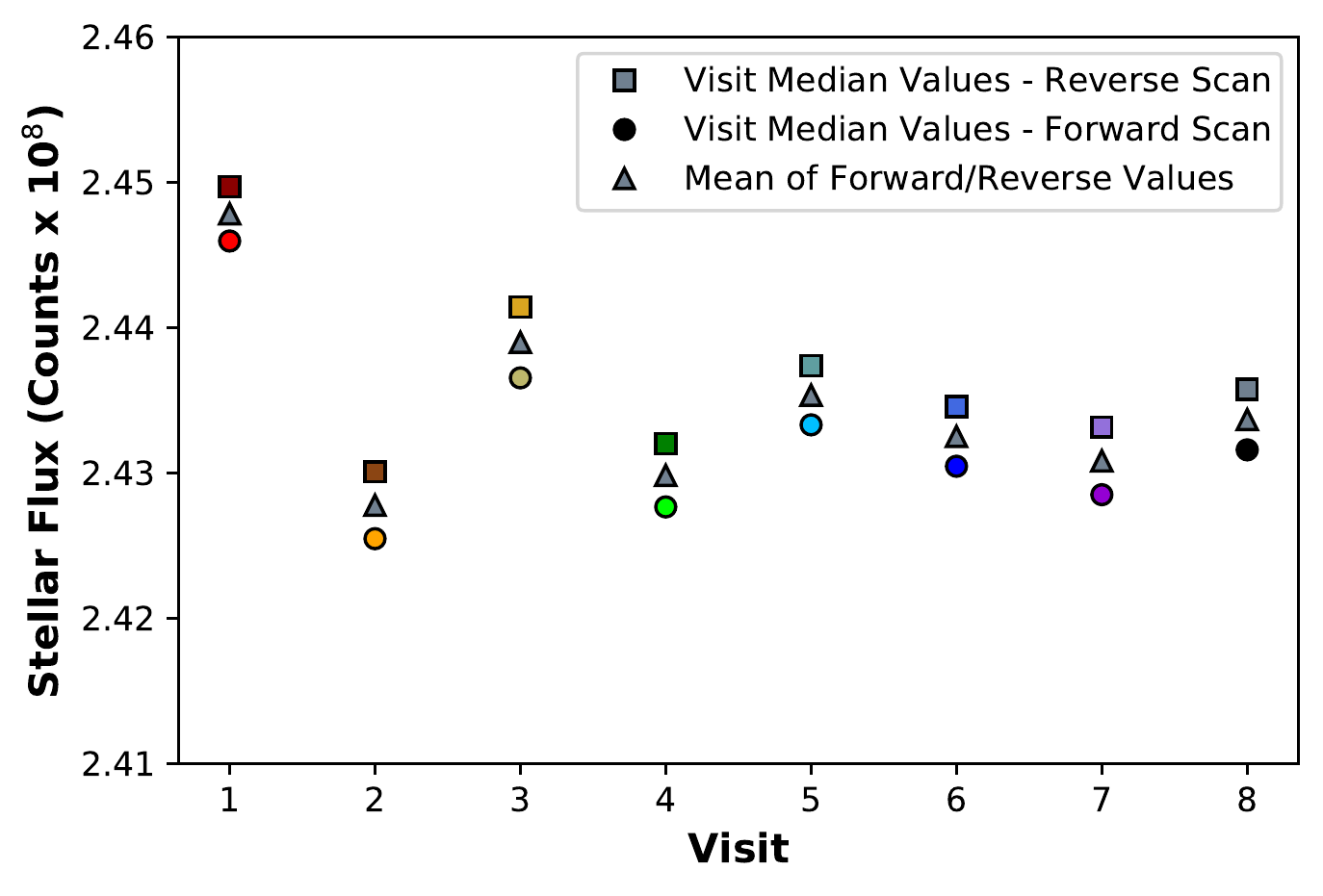}
    \caption{K2-18 varies in average brightness. The HST observations of K2-18 have different average brightness which is possibly due to the different spot coverage fractions during the various observations. However, it is more probable that we are seeing scatter owing to the instrumental limits on repeatability of the WFC3 IR channel.}
    \label{fig:HSTOOT}
\end{figure}

Toward this end, we downloaded all K2-18~b visits from MAST. Following \citet{Tsiaras2019} and \citet{Benneke2019}, we did not analyze the files for visit 06 (taken on 13/05/2018) from proposal 14682 (the ``last" visit chronologically) since this visit suffered from pointing instabilities. We constructed the 2D spectral image for each exposure of the 8 remaining visits using the ``difference read" method described in the Appendix to \citet{Deming2013}, which allows us to easily remove background contamination. We also applied the \citet{Deming2013} methods to remove hot pixels and cosmic rays. For each visit, we constructed a raw white light curve by summing the electron flux in each 2D spectral image and plotting it against phase, where we used the values for T0 and period reported by \citet{Benneke2019} in computing the phase. Since WFC3 data were taken using round-trip spatial scan mode, we analyzed the forward and reverse scan directions separately. To compute the flux from the star, we then found the median out-of-transit (OOT) flux values for each visit, again computing OOT flux values for forward and reverse scan directions separately. For the first 7 visits we analyzed, we used exposures from orbits 2 and 5 to compute the median OOT flux values. For the 8th visit analyzed, we used exposures from orbits 2 and 3 to compute the median OOT flux, since orbit 5 was caught during egress. Finally, we found the mean of the forward and reverse median OOT flux values for each visit. This is the stellar flux value that we report. Figure~\ref{fig:HSTOOT} shows the results of this analysis for all 8 visits. 

The star is brightest during visit 1 when the star is 0.5\% brighter than the average of all visits, and faintest during visit 2 when the star is 0.3\% fainter than the average visit brightness. The peak-to-peak range of 0.8\% approximately matches the maximum range we would expect to see. This is because the variability in the IR should be roughly a factor of two lower than in visible light owing to lower contrast between cooler and hotter regions on the stellar photosphere.

Unfortunately, what is not clear is whether the temporal stability of the flux calibration over two years allows us infer spot coverage from the out-of-transit brightness. Calibration of the WFC3 IR repeatability in the F140W and F160W filters show that 0.5\% (1-$\sigma$) is the best achievable flux stability on timescales comparable to the K2-18 observing baseline \citep{WFC3repeat1, WFC3repeat2}. Therefore, it is probable that the variation we are seeing is caused by instrumental stability noise.


\subsection{The nature of the detection of an atmosphere on K2-18~b}
We have demonstrated stellar contamination is capable of reproducing the signal seen in the HST transmission spectrum of K2-18~b. In a large ensemble of 1000 simulations, a few percent of simulations provide a better fit to the observations than the best-fitting exoplanet model atmosphere, and we find that many synthetic observations of a spotted star would generate a false-positive atmosphere detection. However, we do not rule out the true detection of an exoplanet atmosphere. The choice of whether an atmospheric detection should be preferred over a contaminated spectrum essentially comes down to the prior probability of a spot dominated atmosphere (i.e. which scenario is more likely). There is not a clear winner here. While we know that the star K2-18 has spots and so a contaminated spectrum is not that unlikely, we also expect planets the size of K2-18~b to have low mean-molecular-weight atmospheres detectable with HST/WFC3 \citep{MillerRicci2009,Lopez_2014,Crossfield2017}. Unfortunately, the only conclusive result is that more observations are required to provide confidence in one interpretation over the other. The good news on that front is that K2-18~b looks like an excellent target for characterization with JWST \citep{Changeat2020,Hu2021a,Hu2021b}, and will be observed during Cycle 1 in program GO-2372.

As we move into the JWST era of probing the atmospheres of increasingly smaller planets orbiting small stars (some of the most accessible of which show pronounced activity) with higher precision, the importance of understanding the effects of stellar variability on transmission spectroscopy measurements grows. We briefly explored the impact of spotted stars on JWST observations. We created a suite of simulations of a K2-18~b-like planet transiting an inhomogenous photosphere for two sets of models -- one with 20 large spots and another with 200 small Sun-like spots (with 5\% spot-coverage in each case), and using simulated transmission spectra from the Planetary Spectrum Generator \citep{Villanueva2018} for K2-18~b in the wavelength range from 0.9--5 $\mu$m which covers the range of the JWST NIRISS and NIRCam instruments. 

\begin{figure*}
    \centering
    \includegraphics[width=0.98\textwidth]{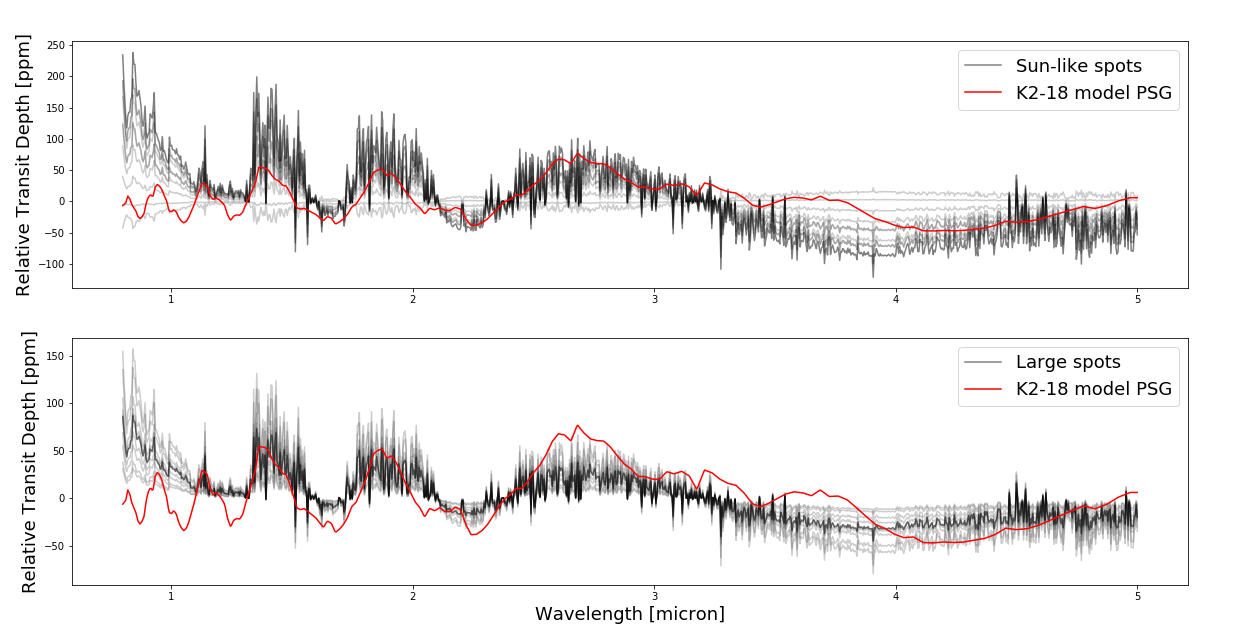}
    \caption{A range of simulated spotted star transmission spectra for a K2-18-like system in the wavelength range covered by JWST's NIRISS and NIRCam instruments (black lines) shows that some regions of the spectrum may produce more reliable detections than others. A model exoplanet atmosphere (in red) is largely similar to the contamination models between 1--3 $\mu$m, but redward of that the models diverge. The upper panel has a model with 200 small, Sun-like spots covering 5\% of the stellar photosphere (black lines), compared to a PSG model for K2-18 (red) while the lower panel has 20 large spots. While most of the spectrum could potentially be corrupted by contamination, the region between 3.5--4 \micron\ could provide less ambiguous detections of atmospheric features.}
    \label{fig:JWST}
\end{figure*}

A range of synthetic contaminated transmission spectra retrieved from this modeling are shown in Figure \ref{fig:JWST}. While stellar contamination between 0.9--2.0 \micron\ causes observations in that region to be unreliable for constraining an exoplanet's atmosphere, contamination seems to be minimal in the 3.5--4 \micron\ region and does not lead to a reduction in signal-to-noise. If a peak is seen in the transmission spectrum at 3.7 $\mu$m then it is more likely to be trustworthy. Additionally, the region between 0.6--1.2 \micron\ looks to be a good diagnostic for determining the level of contamination in the spectrum \citep[this is consistent with the results of ][]{Pinhas2018}. As shown in the figure, observing a high degree of divergence from a flat line in the short-wavelength region of the spectrum can be a signal that there is a spot contamination in the spectrum. It is possible that including additional HST/WFC3 observations in the G102 filter (0.8--1.17 $\mu$m) could provide important diagnostics data that woulds indicate whether contaminated transmission spectra is a concern. Unfortunately, this region is also a diagnostic of hazes in an exoplanet transmission spectrum. Disentangling contamination and hazes may be challenging.

While there is likely no silver bullet to identify potentially contaminated transmission spectra there are a number of methods that show promise. These include observing spectra at high-resolution, so that multiple component stellar models can be fit to the data \citep{Gully2017}, observing at visible and infrared wavelengths simultaneously to measure spot coverage fractions, and collecting long-baseline observation to diagnose changes in the stellar spectrum over time. One area where JWST may benefit is that the very large collecting area allows the detection of atmospheric features with only a single transit. This enables comparisons between transits to be feasible, as compared to the current situation where small planets like K2-18~b require multiple transits with facilities like HST to detect features and do not allow for easy comparison between transits to identify anomalous transits. Even in the JWST-era, claiming and diagnosing a potentially contaminated transmission spectrum is likely going to be made significantly easier by observing multiple transits; any claims of a detection of low signal-to-noise atmospheric features in a single spectrum of a small planet around an active M-dwarf should probably be met with skepticism.



\section{Conclusions}

We presented a stellar contamination model designed to facilitate the interpretation of transmission spectroscopy observations of exoplanets. The model simulates spot-induced stellar variability due to rotation by creating a time-dependent, latitude-longitude, line-of-sight projection of the stellar disk. The disk-integrated contributions from the spot-free photosphere and the spots are used to create the emergent light curve and stellar spectrum as a function of the rotation phase, accounting for the loss of light due to a transiting planet. Finally, the model creates the difference between the out-of-transit and in-transit spectrum to be used as comparison to observations. We applied the model to generate K2-18-like observed transmission spectra, ensuring that in the model is generated physically reasonable star spot models, consistent with the non-contemporaneous optical Kepler observations.
We found that a few percent of random spot model draws provide a better fit to observed HST data than the best-fitting exoplanet model reported by \citet{Benneke2019}. While this paper focuses on the \citet{Benneke2019} analysis, we also examined the \citet{Tsiaras2019} results. While these two data come from the same observations, they run through independent pipelines and do show some differences. However, for our purposes they are functionally the same and the choice of data-set does not impact our conclusions. Additionally, when we create synthetic HST observations from these models, 40\% of observations would result in at least as good a fit to the best-fitting exoplanet atmosphere model as the real observations. Therefore, we have shown that the observations can be well described by a contamination model, and that it is not unlikely that the real observation could have been generated by a spotted star.

We emphasize that while our stellar contamination model is able to produce a water absorption feature, we do not conclude that the signal cannot be the result of actual water absorption from K2-18~b's atmosphere. Our results show that stellar contamination can produce a water absorption feature due to rotationally-modulated variations in the synthetic contaminated transmission spectrum. There are also situations where a bona-fide detection of an exoplanet atmosphere becomes the strongly favored solution such as if spots are very large, or if the stellar variablity is dominated by bright spots rather than cool spots. Further observations are needed to determine the correct model for K2-18~b. Moreover, our model is far from perfect. We do not include effects such as limb darkening, the star rotates as a solid body, and we assume that the spot coverage fraction is persistent over the observations. These additional, more complex effects should be further investigated. New methods, such as those recently published by \citet{Johnson2021} could be used to build on this study. 

K2-18 is an M3V type star with a temperature of 3500 K. This is a particularly problematic area of parameter space in terms of contamination because it is the temperature where water is beginning to become a significant absorber in the spectrum \citep{Rojas2012}. For cooler stars, or crucially here, cooler regions on the photosphere of an M3V star, the water absorption bands are significantly deeper. Such stars are going to be a challenge going forward, even into the JWST era. While the higher precision of JWST may help, we will be probing ever smaller signals and so the issue of stellar contamination of the transmission spectrum is likely to persist. Fortunately, there are various interdisciplinary efforts underway to study the impact of stellar contamination on exoplanet transmission spectra and develop tools and methods to mitigate its impact \citep{Apai2018,Rackham2019b}\footnote{Additionally, NASA's Exoplanet Exploration Program Analysis Group 21: \url{https://exoplanets.nasa.gov/exep/exopag/sag/\#sag21} is focused on understanding stellar contamination in space-based exoplanet transmission spectroscopy.}. New facilities and observations are also being developed with primary goals of quantifying and correcting for stellar contamination, which can be used to reevaluate all existing, and prepare for upcoming, exoplanet transmission spectroscopic observations. NASA has recently selected the Pandora SmallSat \citep{Quintana2021} for further development, which would observe exoplanets and their host stars simultaneously at visible and IR wavelength with long-baseline observations and observe many transits of each star. These data may provide crucial diagnostics of spot coverage levels during observed transits, and provide the insight required to alleviate some of the challenges with stellar contamination. In addition to providing more robust measurements of exoplanet atmosphere features, missions like Pandora will significantly deepen our understanding of stellar photospheric and chromospheric heterogeneities across spectral types.


\acknowledgments
This research is based on observations made with the NASA/ESA Hubble Space Telescope obtained from the Space Telescope Science Institute, which is operated by the Association of Universities for Research in Astronomy, Inc., under NASA contract NAS 5–26555. These observations are associated with program HST-GO-13665 and HST-GO-14682 programs (PI Benneke). This paper includes data collected by the Kepler mission and obtained from the MAST data archive at the Space Telescope Science Institute (STScI). Funding for the Kepler mission is provided by the NASA Science Mission Directorate. STScI is operated by the Association of Universities for Research in Astronomy, Inc., under NASA contract NAS 5–26555.
The material is based upon work supported by NASA under award number 80GSFC21M0002 and by the K2 Guest Investigator Program under award number 80NSSC19K0104. This work was supported by the GSFC Sellers Exoplanet Environments Collaboration (SEEC), which is funded by the NASA Planetary Science Division’s Internal Scientist Funding Mode. DJL was supported by an appointment to the NASA Postdoctoral Program at the NASA Goddard Space Flight Center, administered by Universities Space Research Association under contract with NASA. We thank the anonymous referee for valuable feedback.

\vspace{5mm}
\facilities{
Exoplanet Archive,
HST (WFC3),
Kepler
}

\software{
Cartopy \citep{Cartopy},
Matplotlib \citep{matplotlib},
Numpy \citep{numpy},
Planetary Systems Generator \citep{Villanueva2018}
}



\bibliography{sample63}{}
\bibliographystyle{aasjournal}

\end{document}